\documentclass[pdflatex,iicol,sn-mathphys-num]{sn-jnl}

\usepackage{graphicx}%
\usepackage{amsmath,amssymb,amsfonts}%
\usepackage{amsthm}%
\usepackage{booktabs}%
\usepackage{acro}
\acsetup{single}
\usepackage{bm}
\usepackage[subrefformat=parens,labelformat=parens]{subfig}
\usepackage{makecell}
\usepackage{array}
\usepackage{newtxtext}

\DeclareAcronym{SSS}{
	short=SSS,
	long=single scatter simulation,
}

\DeclareAcronym{DSS}{
	short=DSS,
	long=double scatter simulation,
}

\DeclareAcronym{DL}{
	short=DL,
	long=deep learning,
}

\DeclareAcronym{DLSE}{
	short=DLSE,
	long=deep learning-based scatter estimation,
}

\DeclareAcronym{MRD}{
	short=MRD,
	long=maximum ring difference,
}

\DeclareAcronym{CT}{
	short=CT,
	long=computed tomography,
}

\DeclareAcronym{MR}{
	short=MR,
	long=magnetic resonance,
}

\DeclareAcronym{ET}{
	short=ET,
	long=emission tomography,
}
\DeclareAcronym{PET}{
	short=PET,
	long=positron emission tomography,
}

\DeclareAcronym{AC}{
	short=AC,
	long=attenuation correction,
}

\DeclareAcronym{MC}{
	short=MC,
	long=Monte-Carlo,
}	

\DeclareAcronym{ACF}{
	short=ACF,
	long=attenuation correction factor,
}	

\DeclareAcronym{AF}{
	short=AF,
	long=attenuation factor,
}

\DeclareAcronym{CNN}{
	short=CNN,
	long=convolutional neural network,	
}

\DeclareAcronym{SNR}{
	short=SNR,
	long=signal-to-noise ratio,
}

\DeclareAcronym{TB}{
	short=TB,
	long=total body,
}

\DeclareAcronym{EM}{
	short=EM,
	long=expectation-maximisation,
}

\DeclareAcronym{ML}{
	short=MLEM,
	long=maximum-likelihood,
}	

\DeclareAcronym{MLEM}{
	short=MLEM,
	long=maximum-likelihood expectation-maximisation,
}	

\DeclareAcronym{OSEM}{
	short=OSEM,
	long=ordered-subset \ac{EM},
}

\DeclareAcronym{XCAT}{
	short=XCAT,
	long=extended cardiac-torso,
}

\DeclareAcronym{GPU}{
	short=GPU,
	long=graphical processing unit,
}

\DeclareAcronym{LOR}{
	short=LOR,
	long=line of response,
	long-plural-form = lines of response,
}

\DeclareAcronym{4D}{
	short=4D,
	long=four-dimensional,
}
\DeclareAcronym{3D}{
	short=3D,
	long=three-dimensional,
}

\DeclareAcronym{AI}{
	short=AI,
	long=artificial intelligence,
}

\DeclareAcronym{GGEMS}{
	short=GGEMS,
	long=GPU Geant4-based Monte Carlo simulations,
}

\DeclareAcronym{GATE}{
	short=GATE,
	long=the Geant4 Application for Tomography Emission,
}

\DeclareAcronym{SUV}{
	short=SUV,
	long=standardised uptake value,
}

\DeclareAcronym{mumap}{
	short=${\mu}$-map,
	long=attenuation map,
}

\DeclareAcronym{MSE}{
	short=MSE,
	long=mean squared error,
}

\DeclareAcronym{SRA}{
	short=SRA,
	long=scatter ratio accuracy,
}

\DeclareAcronym{NRMSE}{
	short=NRMSE,
	long=normalised root mean squared error,
}

\DeclareAcronym{GT}{
	short=GT,
	long=ground truth,
}

\DeclareAcronym{SSIM}{
	short=SSIM,
	long=structural similarity index measure,
}

\DeclareAcronym{FOV}{
	short=FOV,
	long=field-of-view,
}

\DeclareAcronym{mmr}{
	short=mMR,
	long={Biograph mMR (SIEMENS Healthineers, Erlangen, Germany)}
}

\DeclareAcronym{DRX}{
	short=DRX,
	long={GE Discovery RX (DRX, GE, Milwaukee, USA)},
}

\DeclareAcronym{FDG}{
	short=[$^{\text{18}}$F]-FDG,
	long=[$^{\text{18}}$F]-fluorodeoxyglucose,
}

\DeclareAcronym{PSMA}{
	short=[$^{\text{18}}$F]-PSMA,
	long=[$^{\text{18}}$F]-prostate-specific membrane antigen-11,
}

\DeclareAcronym{TOF}{
	short=TOF,
	long=time-of-flight
}

\DeclareAcronym{TUM}{
	short=TUM,
	long=Technical University of Munich
}

\DeclareAcronym{LAFOV}{
	short=LAFOV,
	long=long-axial field-of-view
}

\DeclareAcronym{Quadra}{
	short=Vision Quadra,
	long=Siemens Biograph Vision Quadra
}

\DeclareAcronym{ROI}{
	short=ROI,
	long=region of interest,
	long-plural-form = regions of interest
}

\DeclareAcronym{SUVmax}{
	short=SUVmax,
	long=maximum standardised uptake value,
}

\DeclareAcronym{ReLU}{
	short=ReLU,
	long=rectified linear unit,
}

\DeclareAcronym{FWHM}{
	short=FWHM,
	long=full width at half maximum,
}

\DeclareAcronym{BMI}{
	short=BMI,
	long=body mass index,
}

\newcommand{\boldx}{\bm{x}}
\newcommand{\boldy}{\bm{y}}
\newcommand{\bolds}{\bm{s}}
\newcommand{\boldr}{\bm{r}}
\newcommand{\bolda}{\bm{a}}
\newcommand{\boldb}{\bm{b}}
\newcommand{\boldf}{\bm{f}}
\newcommand{\boldu}{\bm{u}}
\newcommand{\boldv}{\bm{v}}
\newcommand{\boldtheta}{\bm{\theta}}

\newcommand{\rbar}{\bar{r}}
\newcommand{\sbar}{\bar{s}}

\newcommand{\boldybar}{\bar{\boldy}}
\newcommand{\boldrbar}{\bar{\boldr}}
\newcommand{\boldsbar}{\bar{\bolds}}
\newcommand{\boldshat}{\hat{\bolds}}

\newcommand{\boldP}{\bm{P}}
\newcommand{\boldA}{\bm{A}}

\newcommand{\calL}{\mathcal{L}}

\newcommand{\R}{\mathbb{R}}

\newcommand{\transp}{^\top}

\newlength{\tempdima}

\newcommand{\rowname}[1]
{\rotatebox{90}{\makebox[\tempdima][c]{\scriptsize#1}}}

\renewcommand{\thesubfigure}{\alph{subfigure}}
\newcommand{\mycaption}[1]
{\refstepcounter{subfigure}(\thesubfigure) {\ignorespaces #1}}

\begin{document}

\title[Article Title]{Evaluation of Deep Learning-based Scatter Correction on a Long-axial Field-of-view PET scanner}

\author[1]{
	\fnm{Baptiste} \sur{Laurent} 
}\email{\texttt{baptiste.laurent@univ-brest.fr}}
\author*[1]{
	\fnm{Alexandre} \sur{Bousse} 
}\email{\texttt{alexandre.bousse@univ-brest.fr}}
\author[1]{
	\fnm{Thibaut} \sur{Merlin} 
}
\author[2]{
	\fnm{Axel} \sur{Rominger} 
}
\author[2]{
	\fnm{Kuangyu} \sur{Shi} 
}
\author[1]{
	\fnm{Dimitris} \sur{Visvikis} 
}

\affil[1]{
	\orgdiv{LaTIM}, \orgname{Inserm UMR 1101, University of Brest}, \city{Brest}, \country{France}
	}
\affil[2]{
	\orgdiv{Dept. Nuclear Medicine, Bern University Hospital,
	University of Bern}, \city{Bern}, \country{Switzerland}
	}

\abstract{
	
	\textbf{Objective:} \Ac{LAFOV} \ac{PET} systems allow higher sensitivity, with an increased number of detected \acp{LOR} induced by a larger angle of acceptance. However, this extended angle increases the number of multiple scatters and the scatter contribution within oblique planes. As scattering affects both quality and quantification of the reconstructed image, it is crucial to correct this effect with more accurate methods than the state-of-the-art \ac{SSS} that can reach its limits with such an extended \ac{FOV}. In this work, which is an extension of our previous assessment of \ac{DLSE} carried out on a conventional \acs{PET} system, we aim to evaluate the \acs{DLSE} method performance on \acs{LAFOV} total-body \acs{PET}.	
	
	\textbf{Approach:} The proposed \acs{DLSE} method based on a \ac{CNN} U-Net architecture uses emission and attenuation sinograms to estimate scatter sinogram. The network was trained from \ac{MC} simulations of \acs{XCAT} phantoms \acs{FDG} PET acquisitions using a \ac{Quadra} scanner model, with multiple morphologies and dose distributions. We firstly evaluated the method performance on simulated data in both sinogram and image domain by comparing it to the \acs{MC} ground truth and \acs{SSS} scatter sinograms. We then tested the method on 7 \acs{FDG} and seven \acs{PSMA} clinical datasets, and compare it to \acs{SSS} estimations.
	
	\textbf{Results:} \Acs{DLSE} showed superior accuracy on phantom data, greater robustness to patient size and dose variations compared to \acs{SSS}, and better lesion contrast recovery. It also yielded promising clinical results, improving lesion contrasts in \acs{FDG} datasets and performing consistently with \acs{PSMA} datasets despite no training with \acs{PSMA}.

	\textbf{Significance:}  \Acs{LAFOV} \acs{PET}  scatter can be accurately estimated from raw  data using the proposed \acs{DLSE} method.
	
}


\acresetall
\maketitle

\section{Introduction}

The emergence of \ac{LAFOV} \ac{PET}/\ac{CT} systems has led to significant advancements in nuclear medicine, providing opportunities to new applications \cite{alberts_lafov_pet,slart_lafov_perspectives, surti_lafov_pet, alberts_lafov_pet2, prenosil_quadra}. These systems provide extended body coverage and improved sensitivity, enabling reductions in either radiopharmaceutical doses or acquisition times \cite{mingels_quadra_high_sensitivity}. However, despite these advantages, the geometry of \ac{LAFOV} \ac{PET} systems is more prone to an increased impact of scatter coincidences due to the larger solid angle of acceptance resulting from the extended \ac{FOV}. This presents challenges for data correction algorithms in terms of both qualitative and quantitative image accuracy.

Scatter coincidences, caused by Compton single or multiple interactions of one or both of the annihilation gamma photons with the attenuation medium, negatively impact quantification and degrade \ac{PET} image quality. Therefore, scatter must be corrected during the reconstruction process. Traditionally, this is achieved through methods such as \ac{SSS} \cite{ollinger_SSS,watson_SSS, watson_new_SSS} and \ac{MC} simulations \cite{barret_MC-SC,holdsworth_MC-SC,levin_MC-SC}. The \ac{SSS} approach, which is the most popular and widely used scatter correction in clinical practice, requires an additional scaling step for the single scatter estimation to account for the presence of multiple scatters. However, since the distribution of multiple scatters differs from that of single scatters \cite{tsoumpas_scatter_2005}, scaling the single scatter distribution cannot accurately account for multiple scatters nor activity from outside the FOV, leading to inaccuracies. To address this,  \ac{DSS} algorithms have been developed, extending the simulation to include second-order scatters \cite{watson_DSS,tsoumpas_scatter_2005}. However, these methods are generally limited to scatter estimation in direct planes or make less accurate estimation of oblique plane scatter contributions, which are much more prominent in \ac{LAFOV} \ac{PET} systems. In addition to variations in scatter distribution along the axial length of the system \cite{Adam_investigation_3Dscatter}, it has been demonstrated that, in \ac{LAFOV} \ac{PET} systems, the scatter distribution also varies with the axial angle of the plane, as does the single-to-multiple scatter ratio \cite{zhang_LAFOV_PET_quantitative_study}. Incorporating oblique planes into scatter estimation can therefore improve the robustness and accuracy of the scatter correction process \cite{bal_3D-SC}.

\Ac{MC}-based methods provide highly accurate scatter estimations \cite{teuho_sss_comparison}, but they are computationally intensive for clinical applications, particularly with \ac{LAFOV} systems, where scatter estimation can take up to an hour per iteration \cite{bayerlein_MC-SC}.

Recent \ac{DL} post-processing techniques aim at correcting scatter in the image domain by transforming uncorrected images into scatter-free ones \cite{jahangir_dl_scatter_correction,guo_dl_scatter_correction}. These methods benefit from shorter computation times and show promising results. However, they disregard important spatial information inherent to \ac{PET} imaging physics, such as Compton scattering, and may introduce artefacts or bias directly into the final image. In addition, as with most \ac{DL}-based image generation approaches, such techniques lack in terms of generalisation and scanner independent applicability.

Within this context, the \ac{DLSE} approach we proposed in \cite{laurent_dlse} estimates  \ac{3D} scatter sinograms from \ac{3D} emission and attenuation sinograms. It has proven effective on conventional \ac{PET} systems, delivering more robust and accurate results than \ac{SSS}. The goal of this study was to evaluate its performance in addressing the specific challenges of scatter correction in \ac{LAFOV} scanners.

The rest of this paper is organised as follows. Section~\ref{section:method} summarises the basics of \ac{PET} image reconstruction with sinogram-based scatter correction using \ac{DLSE}, as well as our evaluation strategy. Section~\ref{sec:results} compares \ac{DLSE} and \ac{SSS} on simulated and patient data and the results are discussed in Section~\ref{sec:discussion}. Section~\ref{sec:conclusion} concludes this work.

\section{Materials and Methods}\label{section:method}

\subsection{Background on Scatter-corrected Image Reconstruction}\label{section:scatter_background}

Image reconstruction consists in estimating the radioactivity distribution image $\boldx = [x_1,\dots,x_J]\transp \in \R^J$ from a measurement sinogram vector $\boldy = [y_1,\dots,y_I]\transp \in \R^I$, with $I$ being the number of \ac{PET} detector bins, and $J$ the number of voxels in the final image. Given an activity distribution $\boldx$, the measurement vector $\boldy$ follows a Poisson distribution with independent entries with expectation
\begin{equation}\label{eq:poisson}
	\mathbb{E} [ \boldy ] = \boldybar(\boldx) 
\end{equation}
where $\boldybar(\boldx)$ is the \ac{PET} forward model, traditionally defined as 
\begin{equation}\label{eq:ybar_vect}
	\boldybar(\boldx) = \tau \boldA \boldP \boldx + \boldrbar + \boldsbar
\end{equation} 
where $\tau$ is the acquisition time, $\boldA = \mathrm{diag}[\bolda]\in\R^{I\times I}$ is a diagonal matrix defined by the \acp{AF} $\bolda = [a_1,\dots ,a_I]\transp \in \R^I$, $\boldP\in\R^{I\times J}$ is the \ac{PET} system matrix defined as $[\boldP]_{i,j} = p_{i,j}$ for all $i,j$,  
$\boldrbar = [\rbar_1,\dots,\rbar_I]\transp\in\R^I$ is the expected random coincidences and $\boldsbar = [\sbar_1,\dots,\sbar_I]\transp\in\R^I$ is the expected scatter. The reconstruction of the image is achieved by finding an image $\boldx$ such that $\boldybar(\boldx) \approx \boldy$ in the sense of $\boldx$ is a maximiser of the Poisson log-likelihood given the measurement sinogram $\boldy$.

The reconstruction is commonly achieved using the \ac{ML}~\ac{EM} algorithm \cite{shepp_OSEM} or its accelerated version \ac{OSEM} \cite{hudson_OSEM}. This require an accurate knowledge of the hyper parameters of $\boldybar$ in \eqref{eq:ybar_vect}: the \acp{AF} $\bolda$, the expected randoms $\boldrbar$ and scatter $\boldsbar$. The \acp{AF} $\bolda$ are computed with a forward projection of the 511-keV attenuation map that is usually derived from a \ac{CT} scan. Randoms can be corrected for by real-time subtraction of a delayed coincidence channel \cite{knoll2010radiation}, and the scatter estimation in clinical systems is widely performed using the SSS approach. \cite{ollinger_SSS,watson_SSS, watson_new_SSS}.

The \ac{DLSE} method we introduced in our previous paper \cite{laurent_dlse} aims to leverage the relations between emission, attenuation and scatter in order to estimate $\boldsbar$ from the measured sinogram data $\boldy$ and \acp{AF} $\bolda$ using a \ac{CNN} $\boldf_{\boldtheta}$. For practical purposes, we exploit the \acp{ACF} $\boldb = [1/a_1,\dots,1/a_I]\transp$ instead of the \acp{AF} $\bolda$. The network $\boldf_{\boldtheta}\colon \R^I \times \R^I \to \R^I$ is then trained such that
\begin{equation}\label{eq:training1}
	\boldf_{\boldtheta} (\boldy , \boldb) \approx  \boldsbar
\end{equation}   
where $\boldtheta$ is the vector of weights to be trained.

In the rest of this paper we assume that the emission data $\boldy$ is random-free, i.e., $\boldrbar = \bm{0}$.

\subsection{DLSE Method}
\subsubsection{Network Architecture}
\label{section:cnn}

\begin{figure}
	\includegraphics[width=\linewidth]{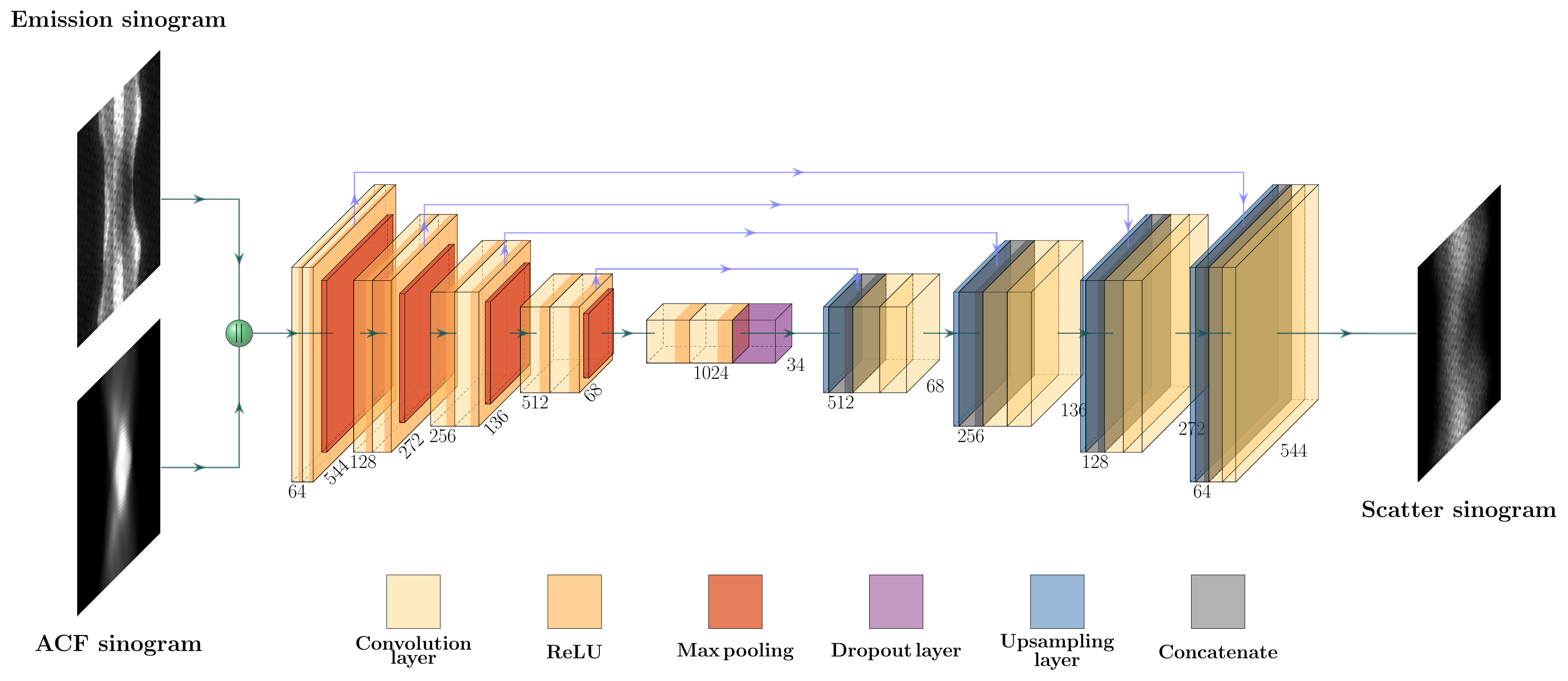}
	\caption{Proposed \ac{DLSE} architecture based on a \ac{CNN} U-Net architecture. The network takes emission and attenuation sinograms as input and predicts the scatter sinogram.}
	\label{fig:unet_architecture}
\end{figure}

The \ac{DLSE} network $\boldf_{\boldtheta}$ is based on a U-Net architecture \cite{ronneberger_UNet}. It processes concatenated \ac{PET} emission (without randoms, as stated in Section~\ref{section:scatter_background}) and \ac{ACF} slices derived from a \ac{3D} sinogram with a dimension of $520 \times 50 \times 11559$ (respectively corresponding to bin displacement, bin angle, and axial slice index) and produces predicted scatter sinogram slices. Each of the five layers of the network contains two $3\times 3$ convolutional layers followed by a \ac{ReLU} activation function and $2\times 2$ max pooling. To prevent overfitting, dropout layers are added at the end of the contraction path. The expansion path is similar to the contraction path but replaces max pooling with $2\times 2$ nearest-neighbour up-sampling to preserve the initial sinogram dimensions. The network ensures that the output dimensions match those of the input sinograms. The architecture is shown in Figure~\ref{fig:unet_architecture}.

The \ac{CNN} was implemented using the Keras framework \cite{chollet2015keras} with a TensorFlow back-end \cite{tensorflow2015-whitepaper}.

\subsubsection{Training}

The data used for training the network consists exclusively of \ac{MC} simulations generated from radiotracer distribution and morphologies. For each experiment, we used a collection of $P$ \ac{MC}-simulated random-free \ac{PET} emission sinograms $\boldy$ and scatter sinograms $\bolds$ as well as the corresponding \acp{ACF} $\boldb$. The training of $\boldtheta$ is achieved via the minimisation problem
\begin{equation}\label{eq:training}
	\min_{\boldtheta} \, \mathbb{E} \left[  \calL \left( \boldf_{\boldtheta} (\boldy, \boldb)  ,  \bolds \right)   \right] 
\end{equation}
where the expectation is taken over the $P$ realisations of  $(\boldy,\boldb,\bolds)$ and  $\calL$ is a loss function we defined as the \ac{MSE}, i.e.,
\begin{equation}\label{eq:mse_training}
	\calL(\boldu,\boldv) = \frac{1}{I} \|\boldu - \boldv\|_2^2 \, ,
\end{equation}
$\|\cdot\|_2$ being the Euclidian norm. 

Note that the forward model  \eqref{eq:ybar_vect} utilises the \emph{expected} scatter sinograms $\boldsbar$ while we train our model with \ac{MC}-simulated sinograms $\bolds$. In principle, the model should be trained with the expected scatter sinograms $\boldsbar = \mathbb{E}[\bolds]$ obtained by averaging several \ac{MC} instances of $\bolds$. However, due to the high computational time, we only used a single instance of $\bolds$ and we assumed $\bolds \approx \boldsbar$. In the following $\bolds$ will be referred to as the \ac{GT}.

The sinograms were normalised for training such that $\boldy$, $\boldb$ and $\bolds$ range between 0 and 1. The output of \ac{DLSE} were then denormalised. We also resized using zero padding so that their dimension is a multiple of $2^5$, five being the number of convolution steps of the network. 

The Adam optimiser \cite{Adam_optimize_Kingma_2014} was used to train the model by solving \eqref{eq:training} for 10 epochs and with a learning rate of $10^{-5}$ as well as a batch size of eight.

\subsubsection{Data Generation} \label{sec-dataset_simulated}

In this section we describe how generated the 3-tuple $(\boldy,\boldb,\bolds)$ used to perform the network training \eqref{eq:training}.

\paragraph{Phantom Generation} 

We used the \ac{XCAT} phantom \cite{segars_XCAT} to produce radiotracer distribution and attenuation maps. As scatters can be affected by object size \cite{ferrero_scatter_object_size}, three different morphologies (Table~\ref{tab-phantom_morphologies}) were generated. \Ac{FDG} activity distributions were then attributed to each organ of the anatomical maps, with distributions inspired from the literature \cite{Lemaitre_simulation}. The activity and associated 511-keV attenuation maps, used to derive the \acp{ACF}, consist of $312 \times 312 \times 200$ images with a $2 \times 2 \times 2$-mm\textsuperscript{3} voxel size. In order to include  scatter originating from outside the scanner's axial \ac{FOV}, the activity along the entire body was considered for the simulation. 
We considered arm-less phantoms  as most \ac{PET} scans are performed in an arms-up position.

Multiple radiotracer doses were used to obtain a representative training dataset of realistic \ac{PET} acquisitions. More specifically, five additional activity distributions were defined by increasing the reference activity by 10\%, 20\% and 30\% as well as reducing it by 10\% and 20\%. For each labelled organ, we added a $\delta$ activity to these new defined distributions, which is randomly set between -5\% and +5\%, to introduce activity variability from organ to organ. These different activity distributions were considered for each of the three different anatomical morphologies.

\begin{table}[!htb]
	\centering
	\begin{tabular}{lrrr} 
		\toprule[\heavyrulewidth]\toprule[\heavyrulewidth]
		& \textbf{Small} & \textbf{Medium} & \textbf{Large}\\  
		\midrule
		Total body height (mm)		& 1,227 & 1,752 & 2,103\\
		Chest short axis (AP) (mm) 	& 163 	& 232 	& 279\\
		Chest Long axis (LAT) (mm) 	& 228  	& 325 	& 391\\
		Chest circumference (mm)	& 696 	& 993 	& 1,194\\
		Waist short axis (AP) (mm)	& 163 	& 233 	& 335\\
		Waist long axis (LAT) (mm)	& 202 	& 288 	& 416\\
		Body weight (kgs)			& 28	& 76	& 129\\
        Body mass index (BMI) ($\mathrm{kg/m^{2}}$) & 18.6 & 24.8 & 29.2\\
		Radiotracer standard dose (MBq)		& 50	& 141 	& 239 \\
		\bottomrule[\heavyrulewidth] 
	\end{tabular}
	\caption{Anatomical characteristics of the three simulated phantom morphologies.}
	\label{tab-phantom_morphologies}
\end{table}

\paragraph{Monte-Carlo Simulations}

\Ac{MC} tools have been shown to accurately simulate the Siemens Vision 600 and the Vision Quadra scanner models, as evidenced by their good agreement with experimental data \cite{penaacosta_2024,salvadori_2024}. The Vision Quadra model geometry within \ac{GATE} software \cite{Gate_Jan_2004} consists of four Siemens Vision 600 units aligned with an axial gap between them.

We performed \ac{MC} simulations using the \ac{XCAT}-generated phantoms and the \ac{GATE} software  to simulate data acquisition on the \ac{Quadra} \ac{PET} scanner \cite{prenosil_quadra} (characteristics shown in Table~\ref{tab-param-simu}). The simulations were handled using back-to-back gamma photons sources, generated from the three distinct morphologies and six varied activity distributions, as described in the previous section, resulting in a total of 18 simulations.
\begin{table}
	\centering
	\begin{tabular}{rl} 
		
		\toprule[\heavyrulewidth]\toprule[\heavyrulewidth]
		Ring diameter &  82~cm \\
		Transaxial \ac{FOV}	&  726~mm	\\
		Axial \ac{FOV} & 106~cm \\
		Crystal material &  LSO \\
		Crystal dimensions &  3.2$\times$3.2$\times$20~mm \\
		Total number of crystals &   243,200 \\
        Crystal Rings   & 320 \\
		Crystals per ring &  760 \\
		Crystals per detector block & 64 (8$\times$8) \\
		Energy window &  435--585~keV \\
		Coincidence window &  4.7~ns\\
		\bottomrule[\heavyrulewidth]
	\end{tabular}
	\caption{Siemens Vision Quadra characteristics \cite{prenosil_quadra}.}
	\label{tab-param-simu}
\end{table}

The sinograms were  generated considering the full-angle acceptance mode of the Vision Quadra, with a \ac{MRD} of 322. The number of coincidences of the simulated measurement sinograms varied from $0.44\times10^9$ to $2.40\times10^9$, with a mean of $1.2\pm 0.55 \times 10^9$, with a mean scatter ratio of 31.5\% with a minimum of 26.7\% (for the small phantom with a \ac{BMI} of 18.6) up to 34.6\% (for the large phantom with a \ac{BMI} of 29.9). These scatter fractions as a function of \ac{BMI} correspond well with those shown previously in the literature \cite{ghabrial_lafov_scatter_fraction}. The single scatters contribution reached a ratio of 28.1\% (from 24.4\% to 30.5\%) while the mean multiple scatters reached 3.4\% (from 2.2\% to 4.2\%).

Each activity, scatter and \ac{ACF} sinogram consists of 11,559 sinograms slices (with dimensions of $520 \times 50$), which represents a total of $P=208,062$ realisations of $(\boldy,\boldb,\bolds)$, allocated as follows: 2/3 for training, 1/6 for validation and 1/6 for testing.

\subsection{Evaluation}

\subsubsection{Simulated Data}\label{section:simulated_data}
The phantoms used for evaluation were not included in the training.
\Ac{DLSE} scatter sinograms were compared to the \ac{MC} \ac{GT} and \ac{SSS} scatter sinograms. In addition to visualising scatter distributions and profile lines, we used \ac{NRMSE} to evaluate its robustness according to patient size and injected dose variations, with \ac{NRMSE} defined as
\begin{equation}\label{eq:NRMSE}
	\mathrm{NRMSE}(\boldshat,\bolds) = \frac{\calL(\boldshat,\bolds)  }{\boldshat_{\max} - \boldshat_{\min}}
\end{equation}
where $\bolds$ is the \ac{MC} \ac{GT} sinogram and $\boldshat$ is the estimated one (by either \ac{DLSE} or \ac{SSS}),  $\boldshat_{\max}$ and $\boldshat_{\min}$ being respectfully the maximal and minimal values of $\boldshat$, and $\calL$ is the \ac{MSE} loss defined in \eqref{eq:mse_training}.

We then performed an evaluation in the image domain, after a reconstruction of the sinograms with \ac{OSEM} (3 iterations, 24 subsets) followed by post-reconstruction filtering (4-mm-\ac{FWHM} Gaussian filter) using Siemens e7tools toolkit. Images were reconstructed into a $440\times440\times645$ matrix with 1.65-mm cubic voxels. \Ac{MC} \ac{GT} and \ac{DLSE} sinograms were smoothed by a 2-pixel-\ac{FWHM} Gaussian filter before being injected into the forward model $\boldybar$ used for reconstruction (i.e., the expected scatter $\boldsbar$ in \eqref{eq:ybar_vect}).

We evaluated the \ac{DLSE} performance in various organs, by defining four 50-mm-diameter spherical \acp{ROI}, in which we compute the mean \ac{SUV} activity error.  
We finally simulated a phantom incorporating  six spherical lesions  of various sizes (10~mm, 20~mm and 40~mm) and locations (three in the lungs, three in the liver) to assess the \ac{DLSE} performance when faced with cold or hot abnormal activities. The simulation was performed for three lesion contrasts, 0:1, 3:1 and 6:1, where the  contrast is computed as
\begin{equation}\label{eq:lesion_contrast}
	\text{lesion contrast}=\frac{\text{mean lesion activity}}{\text{mean organ activity}}
\end{equation}

\subsubsection{Clinical Data}

In a first step, we used seven \ac{FDG} clinical datasets acquired with a Vision Quadra total body scanner to assess the \ac{DLSE} performance on real clinical data. The 10-minute acquisitions, with patient weights ranging from 52 to 98~kgs and mean injected dose ranging from 160 to 297~MBq, leaded to a detected coincidences count varying from $2.5\times10^9$ to $4.7\times10^9$. The data were reconstructed using the same process as described in Section~\ref{section:simulated_data}. We subsequently compared mean activities within multiple regions of interest of \ac{DLSE} and \ac{SSS} corrected \ac{PET} images.

In addition, we applied our \ac{FDG}-trained \ac{DLSE} model to seven \ac{PSMA} clinical datasets. The patients' mean weights ranged from 51 to 98kg, with mean injected doses between 197 and 204~MBq. The four minutes acquisitions provided from $579\times10^6$ up to $1~038\times10^6$ detected coincidences. This assessment aimed to evaluate the capability of the \ac{DLSE} approach to generalise across various radiopharmaceuticals beyond those included in its training. Our analysis focused particularly on prostate lesions and associated metastases, comparing the resulting activity levels in \ac{PET} images corrected by \ac{DLSE} and \ac{SSS}. 

The characteristics of the acquisition are summarised in Table~\ref{tab-donnees_geometriques_scans}.
The emission sinograms were corrected for randoms before running  \ac{DLSE}.

\begin{table}
	\centering
	\begin{tabular}{rcc} 
		
		\toprule[\heavyrulewidth]
								& \acs{FDG} 			& \acs{PSMA} \\ 
		\midrule
		Patient weight (kgs)	& $77.6\pm15.1$		& $74.4\pm13.6$ \\
        Body mass index (BMI) ($\mathrm{kg/m^{2}}$)   & $26.6\pm6.38$     & $25.0\pm3.52$  \\
		Injected activity (MBq)	& $233.4\pm45.5$ 	& $199.7\pm2.4$ \\
		Coincidences ($\times10^9$) & $3.6\pm0.6$	& $0.8\pm0.2$	\\
		\bottomrule[\heavyrulewidth]
	\end{tabular}
	\caption{Clinical acquisition characteristics.}
	\label{tab-donnees_geometriques_scans}
\end{table}

\section{Results}\label{sec:results}

\subsection{Simulated Data}

In the following results reporting, the metrics are averaged over the testing dataset. For example, the \ac{NRMSE} on the small phantom is averaged over all doses, while the \ac{NRMSE} for a given dose is averaged over all phantom sizes.

\subsubsection{Scatter Sinograms Estimation}

In this section we compare the performances of \ac{DLSE} and \ac{SSS} for the estimation of $\bolds$. The \ac{NRMSE} is computed between the estimated scatter $\boldshat$ and the \ac{GT} scatter $\bolds$.

Figure~\ref{fig:sino_simulated} shows the emission sinogram $\boldy$, the \acp{ACF} $\boldb$, the \ac{GT} scatter $\bolds$, and the outputs of \ac{DLSE} and \ac{SSS} (scatter sinograms were smoothed using a 4-mm \ac{FWHM} Gaussian filter) under two conditions: (i) a low-dose setting (standard dose $-20\%$, see Table~\ref{tab-phantom_morphologies} for the standard dose) with a small phantom, and (ii) a high-dose setting (standard dose $+30\%$) with a large phantom. The \ac{DLSE} scatter sinogram distribution resembles qualitatively the \ac{GT} in both settings, with no visible artefacts. In the small phantom scenario, the scatter profiles show that the \ac{DLSE} scatter distribution aligns closely with the \ac{GT} on the tails, although the activity peak is slightly underestimated. In contrast, \ac{SSS} exhibits some errors on the tails but accurately matches the maximum scattered activity. On the other hand, in the large phantom example, \ac{SSS} underestimates scatter activity, whereas \ac{DLSE} more closely approximates the \ac{GT}.

\newcommand\widthfactor{0.99}
\begin{figure*}

	\centering
	
	\begin{tabular}{m{.1\textwidth} m{.4\textwidth}m{.4\textwidth}}
		& \makecell{Small phantom\\Standard dose $-20\%$}  & \makecell{Large phantom\\Standard dose $+30\%$ }\\
		 Emission  & 
		\includegraphics[width=\widthfactor\linewidth]{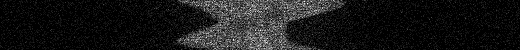} &
		\includegraphics[width=\widthfactor\linewidth]{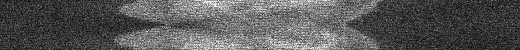} \\
		
		\scriptsize ACF & 
		\includegraphics[width=\widthfactor\linewidth]{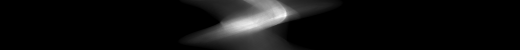} &
		\includegraphics[width=\widthfactor\linewidth]{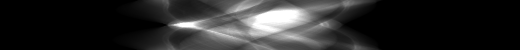} \\
		
		\scriptsize GT scatter & 
		\includegraphics[width=\widthfactor\linewidth]{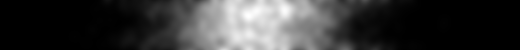} &
		\includegraphics[width=\widthfactor\linewidth]{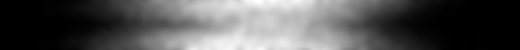} \\
		
		\scriptsize DLSE & 
		\includegraphics[width=\widthfactor\linewidth]{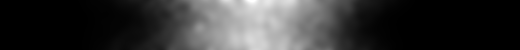} &
		\includegraphics[width=\widthfactor\linewidth]{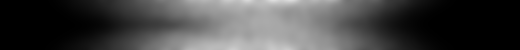} \\
		
		\scriptsize SSS & 
		\includegraphics[width=\widthfactor\linewidth]{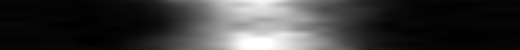} &
		\includegraphics[width=\widthfactor\linewidth]{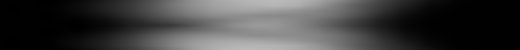} \\
		
		\scriptsize Profiles & 
		\includegraphics[width=\widthfactor\linewidth]{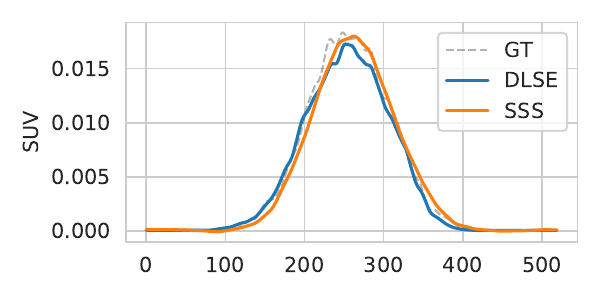} &
		\includegraphics[width=\widthfactor\linewidth]{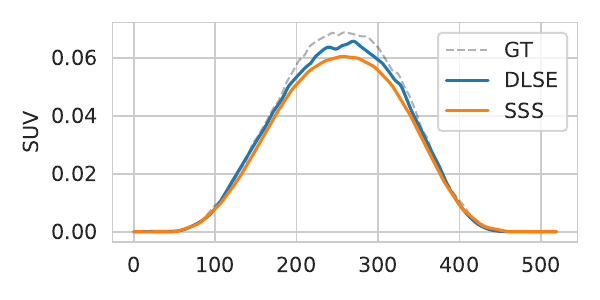} \\
		
	\end{tabular}	
	\caption{
		Input activity and attenuation sinograms, followed by \ac{GT}, \ac{DLSE} and \ac{SSS} scatter sinograms as well as their profile lines along the bins axis. The small and large phantom datasets correspond  to sinogram slices from the neck and lungs respectively. The profiles were obtained by averaging the sinograms over all azimuthal angles.
	}
	\label{fig:sino_simulated}
\end{figure*}

\begin{figure}
	\centering
	\subfloat[]{
		\includegraphics[width=0.50\linewidth]{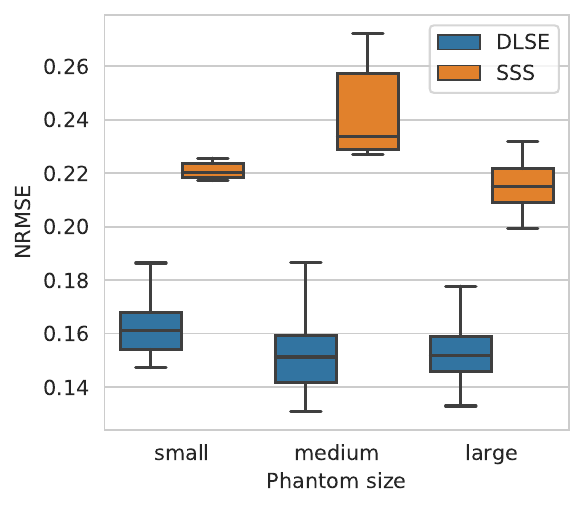}
		\label{subfig:sinos_phantom_size}}
	\subfloat[]{
		\includegraphics[width=0.50\linewidth]{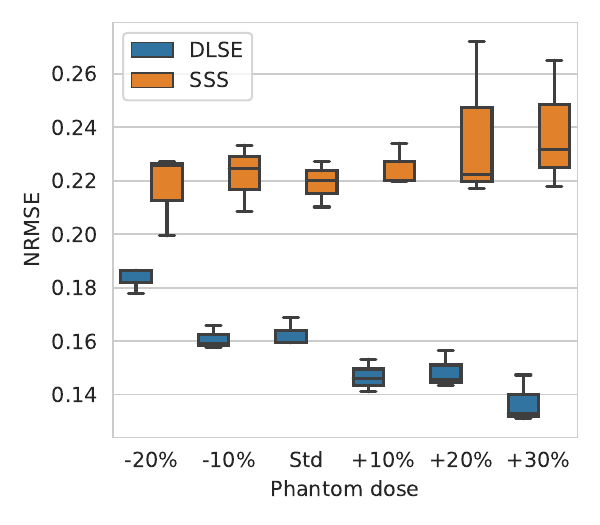}
		\label{subfig:sinos_phantom_dose}}
	\caption{ 
		\Ac{NRMSE} of the \Ac{DLSE}- and \ac{SSS}-estimated sinograms according to \protect\subref{subfig:sinos_phantom_size} the phantoms size and  \protect\subref{subfig:sinos_phantom_dose} the dose. 
		 }
	\label{fig:sinos_simulated_results}
\end{figure}

Figure~\ref{fig:sinos_simulated_results} presents the  \ac{NRMSE} between the \ac{GT} scatter $\bolds$ and the estimated scatter $\boldshat$ across various phantom sizes and doses. We observe that \ac{DLSE} outperforms \ac{SSS} in both experiments. \Ac{DLSE} appears to be less sensitive to the phantom size than \ac{SSS}, with  \ac{NRMSE} ranging from $0.153 \pm 0.016$ to $0.163 \pm 0.014$ for \ac{DLSE} and from $0.215 \pm 0.012$ to $0.243 \pm 0.020$ for \ac{SSS}. We also note that the performance of \ac{DLSE} improves gradually with increasing injected dose, with  \ac{NRMSE} ranging from $0.184 \pm 0.005$ to $0.137 \pm 0.009$. In contrast, the accuracy of \ac{SSS} slightly decreases with increasing injected dose, where the  \ac{NRMSE} ranges from $0.217 \pm 0.016$ to $0.238 \pm 0.024$.

\subsubsection{Reconstructed Images}

The next analysis is conducted on reconstructed images from \ac{MC} simulated data $\boldy$ using the reconstruction process described in Section \ref{section:simulated_data} and estimated scatter $\boldshat$ obtained from \ac{DLSE} and \ac{SSS}.

\begin{figure*}

	\settoheight{\tempdima}{\includegraphics[height=0.15\linewidth]{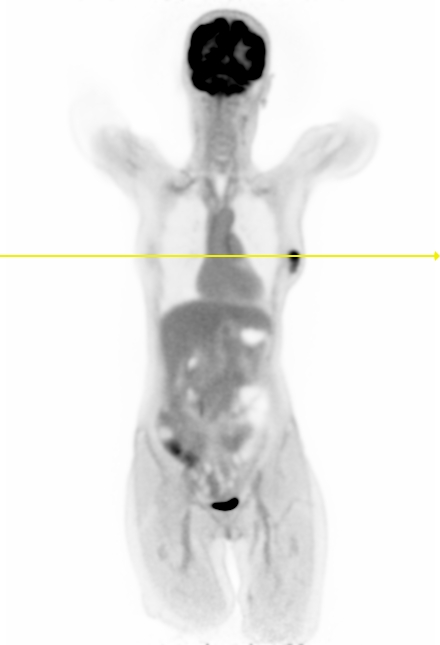}}%
	\centering
	
	\begin{tabular}{c@{\hspace{-0.cm}}c@{\hspace{-0.35cm}}c@{\hspace{-0.35cm}}c@{\hspace{-0.35cm}}c@{}c} 
		& \scriptsize No correction  & \scriptsize Reference & \scriptsize DLSE & \scriptsize SSS & \scriptsize Profiles \\
		
		\rowname{\makecell{Lesion contrast 6:1\\10mm liver lesion}} & 
		\includegraphics[width=\tempdima]{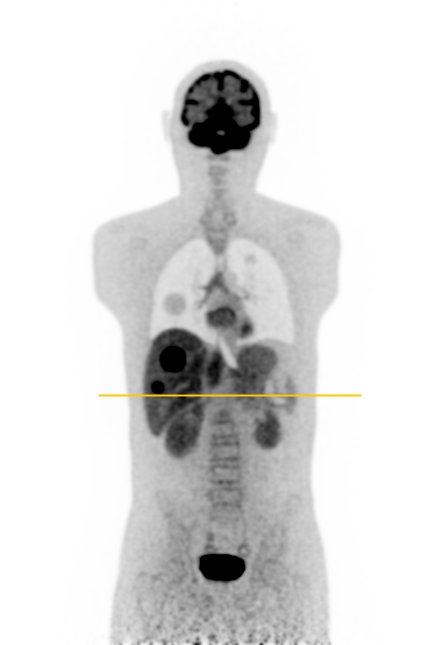} &
		\includegraphics[width=\tempdima]{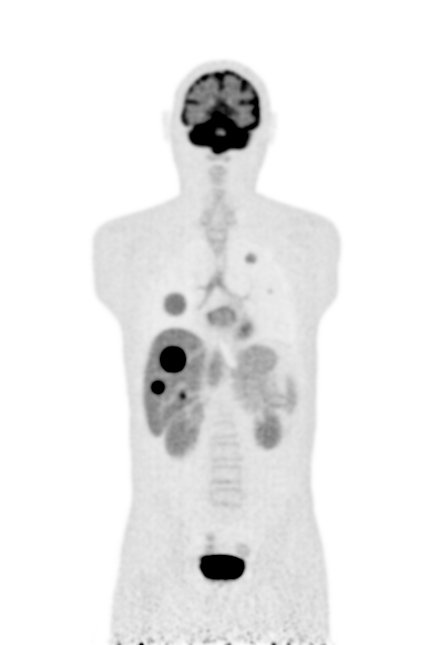} &
		\includegraphics[width=\tempdima]{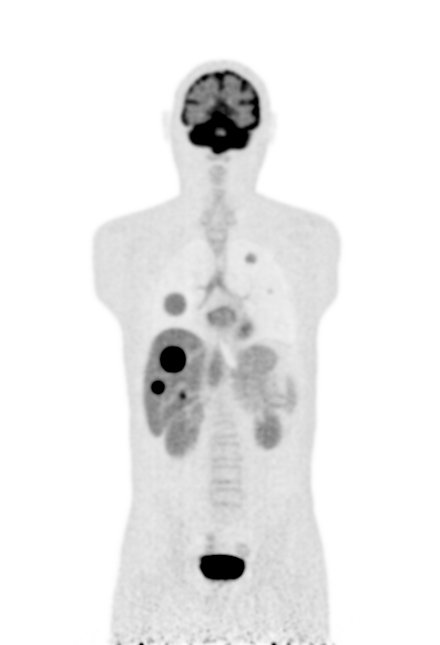} &
		\includegraphics[width=\tempdima]{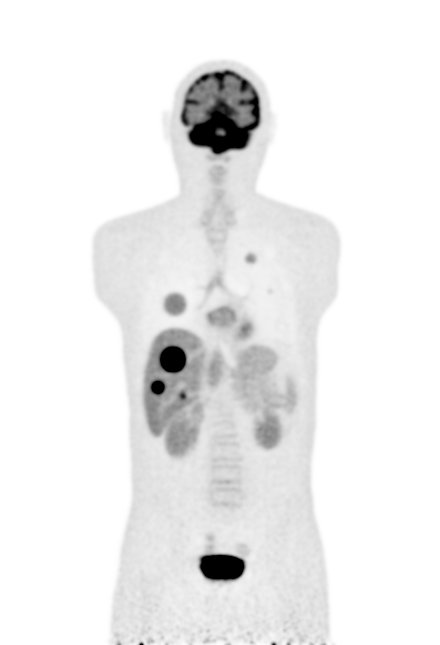}&
		\includegraphics[width=1.6\tempdima]{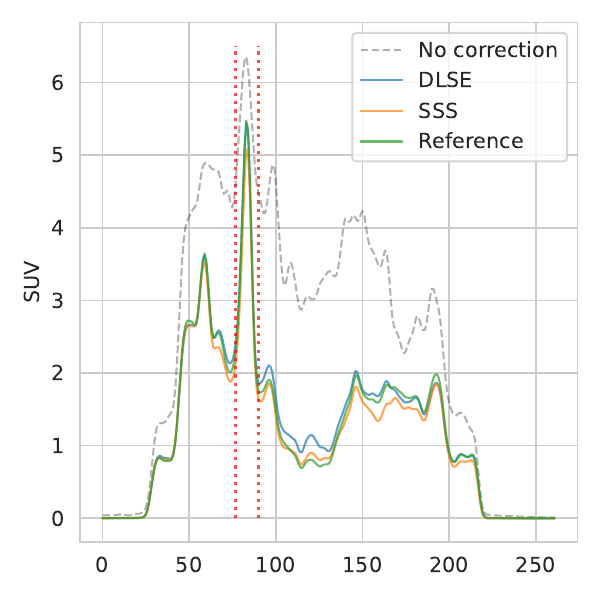} \\

		\rowname{\makecell{Lesion contrast 3:1\\20mm lung lesion}} & 
		\includegraphics[width=\tempdima]{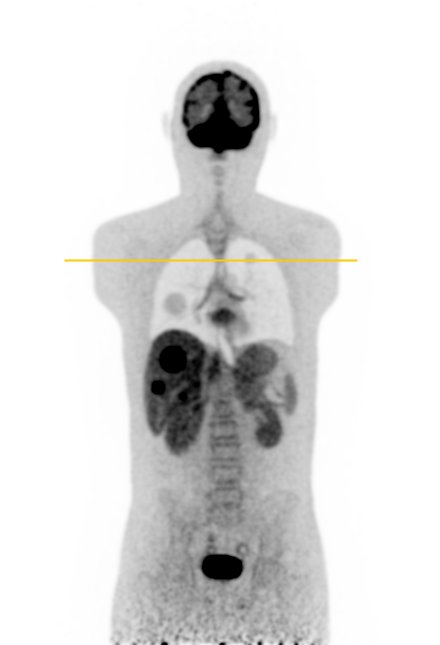} &
		\includegraphics[width=\tempdima]{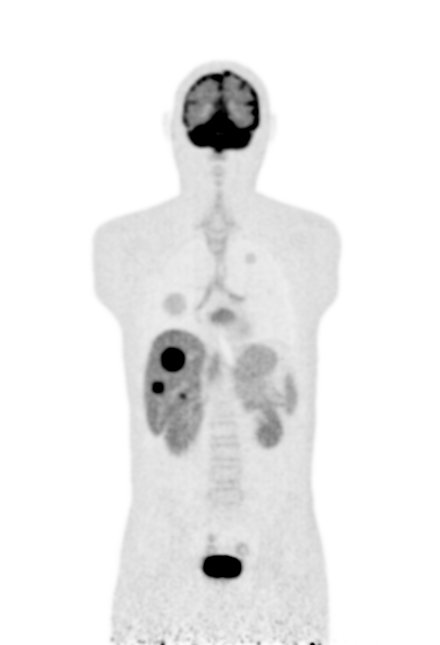} &
		\includegraphics[width=\tempdima]{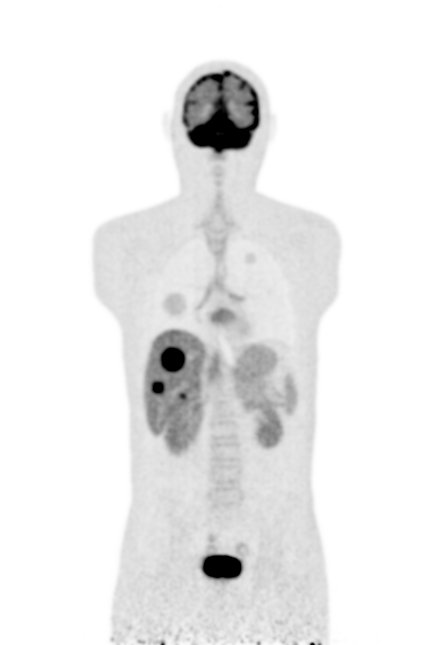} &
		\includegraphics[width=\tempdima]{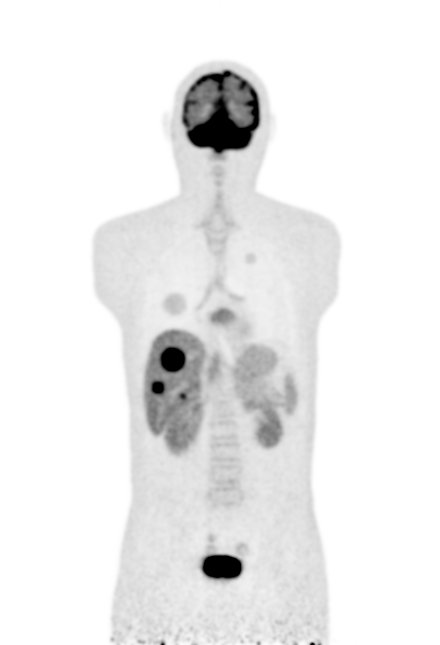}&
		\includegraphics[width=1.6\tempdima]{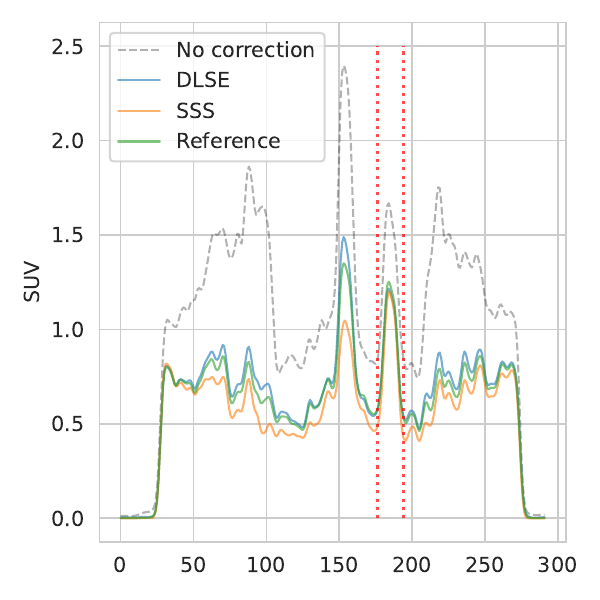} \\

		\rowname{\makecell{Lesion contrast 0:1\\40mm lung lesion}} & 
		\includegraphics[width=\tempdima]{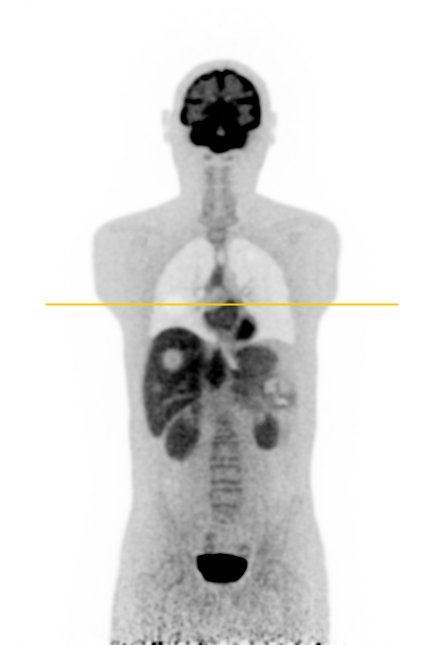} &
		\includegraphics[width=\tempdima]{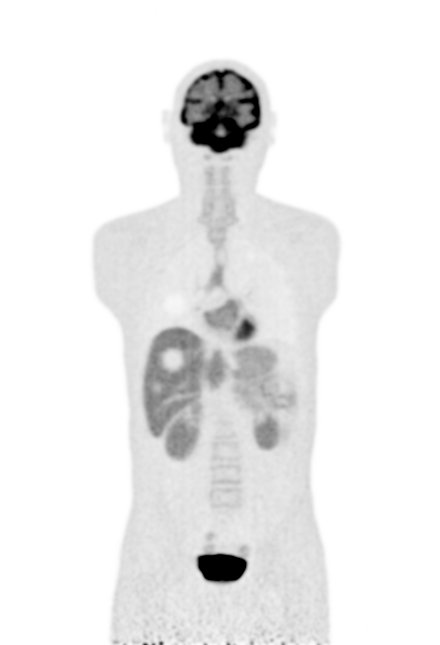} &
		\includegraphics[width=\tempdima]{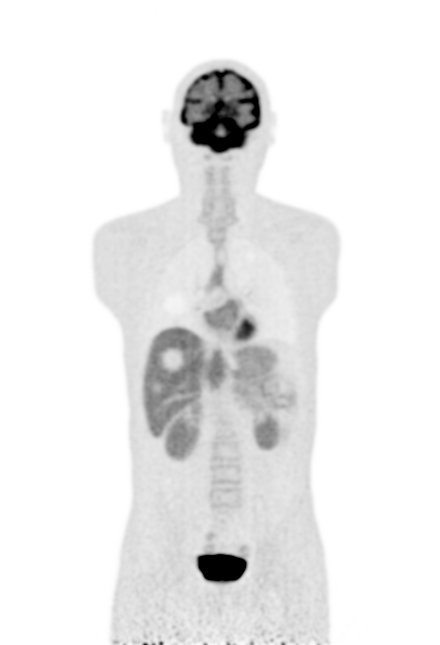} &
		\includegraphics[width=\tempdima]{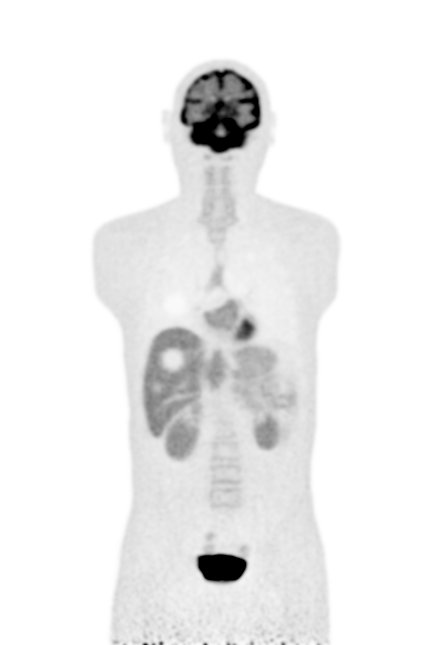}&
		\includegraphics[width=1.6\tempdima]{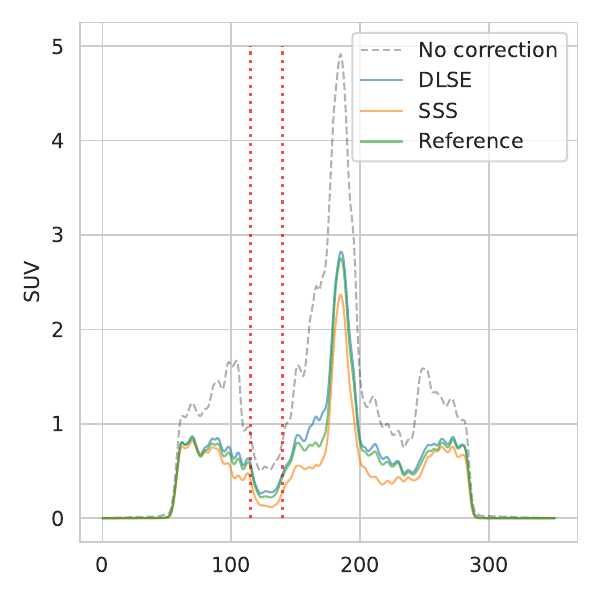} \\

	\end{tabular}	
	\caption{
		 Reconstructed images using simulated data without and with scatter correction  using \ac{SSS} and \ac{DLSE} corrections. The reference image corresponds to the reconstruction from scatter-free data. Profiles are shown along the yellow lines drawn of the first column. Lesion borders are represented with the vertical dashed red lines.
		 }
	\label{fig:images_simulated_lesions}
\end{figure*}

Figure~\ref{fig:images_simulated_lesions} shows the reconstructed images with and without scatter correction. The image reconstructed from scatter-free data is referred to as the ``reference image''. The \ac{DLSE}-scatter corrected images are visually similar to the \ac{MC} \ac{GT} and display no artefacts. The profile analysis demonstrates that \ac{DLSE}-corrected images yield more accurate estimates of activity in the 10-mm liver lesion with a 6:1 contrast (first row). In this case, \ac{SSS} correction underestimates the lesion activity due to an overestimation of scatter contribution within the region. Overall, the \ac{DLSE} profile line aligns more closely with the reference image compared to the \ac{SSS} correction. In the profile lines of the second row, representing a 20-mm lung lesion with a 3:1 contrast, both correction methods lead to an accurate estimate of the lesion \ac{SUVmax}. However, the \ac{SSS} correction underestimates activity in the lung region, particularly in the trachea. A similar pattern is observed in the profile lines of the final row, which depict a necrotic 40-mm lung lesion (0:1 contrast). \Ac{SSS} correction tends to over-correct for scatter in lungs and trachea, although it performs well in the shoulder region. The \ac{DLSE} profile line shows good agreement with the reference image but exhibits an activity overestimation in the area between the lesion and the heart.

\begin{figure}
	\centering
	\subfloat[]{
		\includegraphics[width=0.50\linewidth]{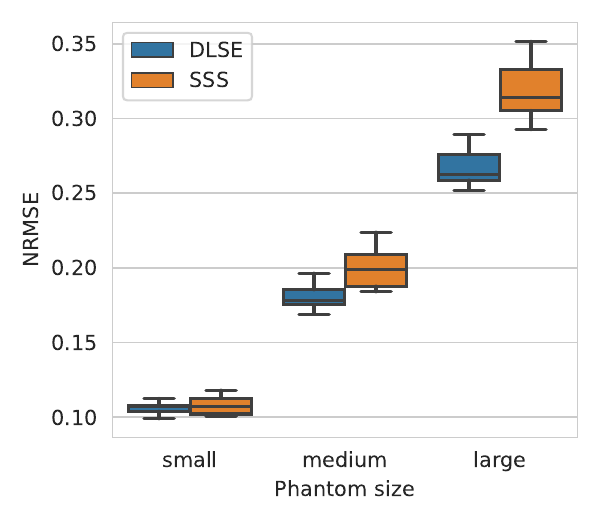}
		\label{subfig:images_phantom_size}}
	\subfloat[]{
		\includegraphics[width=0.50\linewidth]{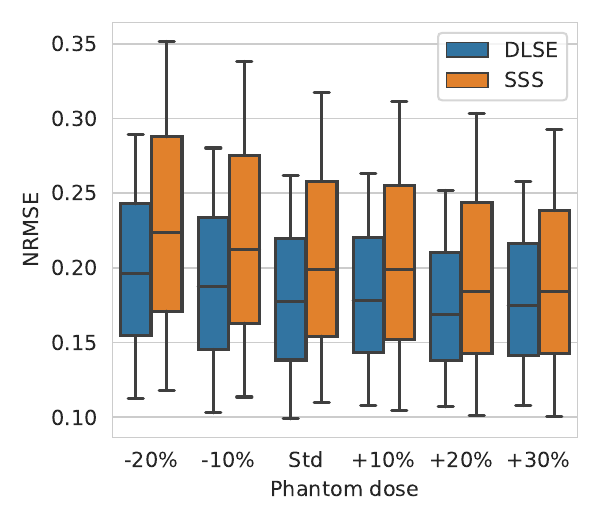}
		\label{subfig:images_phantom_dose}}
	\caption{ \Ac{NRMSE} of the \ac{DLSE}- and \ac{SSS}-reconstructed images  according to \protect\subref{subfig:images_phantom_size} the phantom size and \protect\subref{subfig:images_phantom_dose} the dose.
	}
	\label{fig:simulated_images_results}
\end{figure}

Figure~\ref{fig:simulated_images_results} shows the \ac{NRMSE} results for \ac{DLSE}- and \ac{SSS}-corrected reconstructed images, computed using the scatter-free reconstructed image as a reference. We observe that performance decreases for both methods as the phantom size increases. \Ac{DLSE} outperforms \ac{SSS} across all phantom sizes, with \ac{NRMSE} ranging from $0.106 \pm 0.005$ to $0.267 \pm 0.014$ for \ac{DLSE} and from $0.108 \pm 0.007$ to $0.319 \pm 0.022$ for \ac{SSS}. We also observe that performance improves with the injected dose for both methods, with \ac{DLSE} again outperforming \ac{SSS} at all dose levels, confirming the results shown in Figure~\ref{subfig:sinos_phantom_dose}. For \ac{DLSE}, the \ac{NRMSE} decreases from $0.199 \pm 0.088$ to $0.176 \pm 0.073$, while for \ac{SSS}, it decreases from $0.231 \pm 0.117$ to $0.192 \pm 0.096$.

\begin{figure}
	\centering	
	\subfloat[Mean activity errors within different \acp{ROI}]{
		\includegraphics[width=0.5\linewidth]{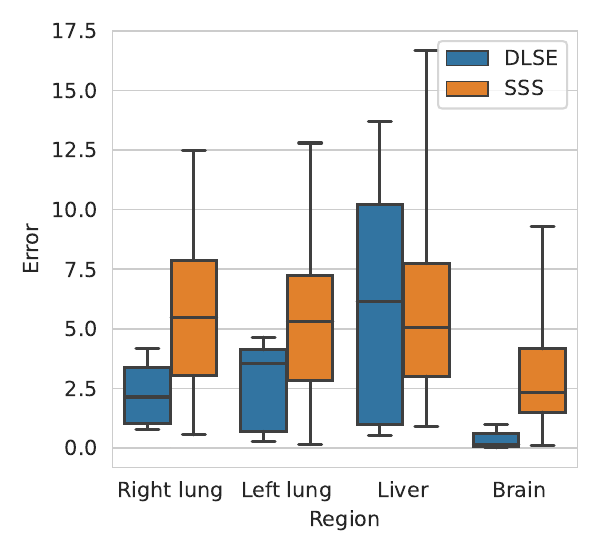}
		\label{subfig:images_simulated_ROI}}
	\subfloat[Lesion contrast]{
		\includegraphics[width=0.5\linewidth]{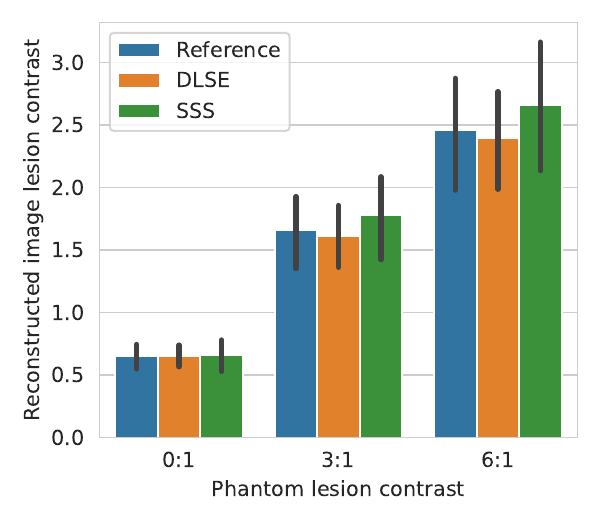}
		\label{subfig:images_phantom_lesions_means}}
	\caption{
		\Ac{DLSE} and \ac{SSS} reconstructed images results for different \protect\subref{subfig:images_simulated_ROI} \acp{ROI} and \protect\subref{subfig:images_phantom_lesions_means} lesion contrasts. 
		}
	\label{fig:simulated_lesions}
\end{figure}

Figure~\ref{fig:simulated_lesions} shows quantitative results in the reconstructed images over different \acp{ROI}. Figure~\ref{subfig:images_simulated_ROI} shows the \ac{NRMSE} on specific organs. While both \ac{DLSE} and \ac{SSS} methods shows similar \ac{NRMSE} on liver region ( $6.30 \pm 4.74$ versus $7.09 \pm 6.53$), \ac{DLSE} outperforms \ac{SSS} on lungs ($2.21 \pm 1.21$ versus $5.66 \pm 3.40$ for right lung and $2.83 \pm 1.72$ versus $5.23 \pm 3.30$ for left lung), and brain regions ($0.31 \pm 0.34$ versus $3.40 \pm 3.38$). Figure~\ref{subfig:images_phantom_lesions_means} shows the contrast on the three lesions (0:1, 3:1 and 6:1) for the reference image, and \ac{DLSE} \ac{SSS}-corrected images, computed following \eqref{eq:lesion_contrast}. \Ac{DLSE} yields closer lesion contrasts to the reference image than \ac{SSS}, regardless of the simulated contrast. For necrosed lesions, with a contrast of 0:1, the mean lesion contrasts in the reconstructed image are respectively, $0.649$, $0.657$ and $0.649$ for \ac{DLSE}, \ac{SSS} and reference methods. At a 3:1 contrast, the values are $1.611$, $1.778$, and $1.657$, and at a 6:1 contrast, they are $2.400$, $2.657$, and $2.459$, respectively.

\subsection{Clinical Data}

\subsubsection{FDG Datasets}

\begin{figure*}

		\settoheight{\tempdima}{\includegraphics[height=0.15\linewidth]{FDG_038_noSC.jpg}}%
		\centering
		
		\begin{tabular}{@{\hspace{-0.22cm}}c@{\hspace{-0.25cm}}c@{\hspace{-0.25cm}}c@{\hspace{0cm}}c@{}} 
				\scriptsize No correction  & \scriptsize DLSE & \scriptsize SSS & \scriptsize Profiles \\
				\includegraphics[width=\tempdima]{FDG_038_noSC.jpg} &
				\includegraphics[width=\tempdima]{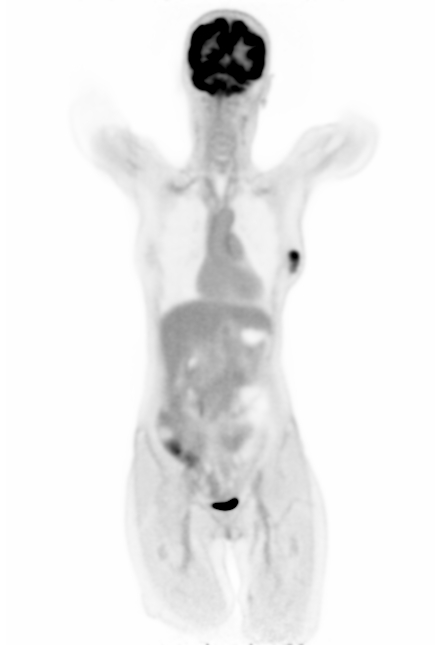} &
				\includegraphics[width=\tempdima]{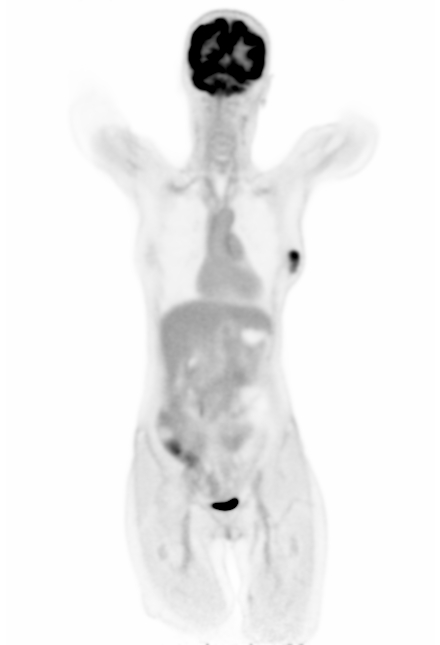} &
				\includegraphics[width=2\tempdima]{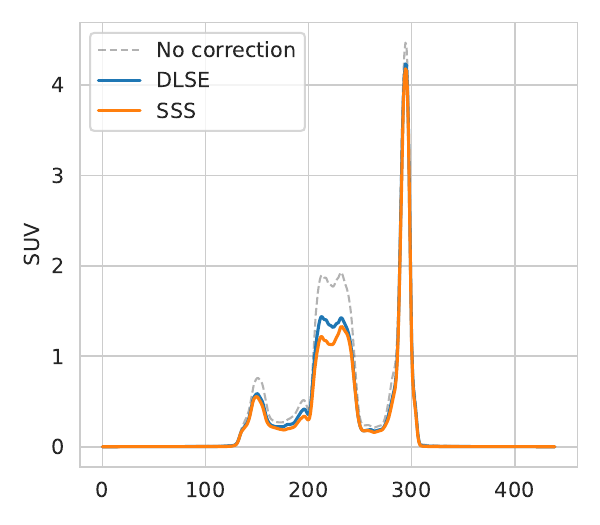} \\
				
				\includegraphics[width=\tempdima]{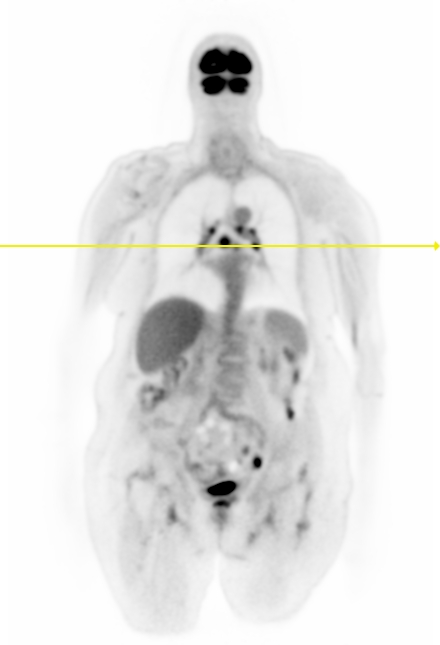} &
				\includegraphics[width=\tempdima]{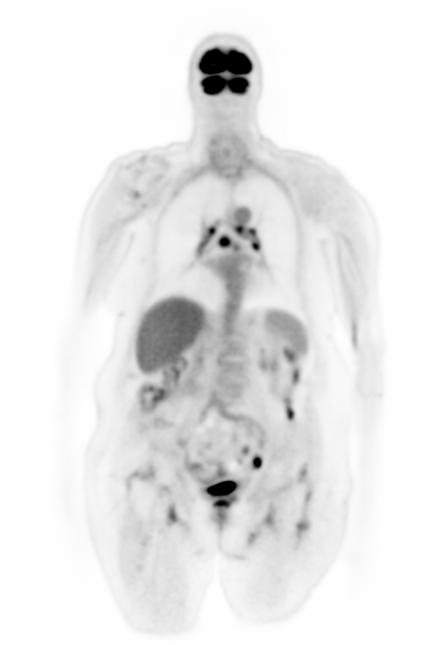} &
				\includegraphics[width=\tempdima]{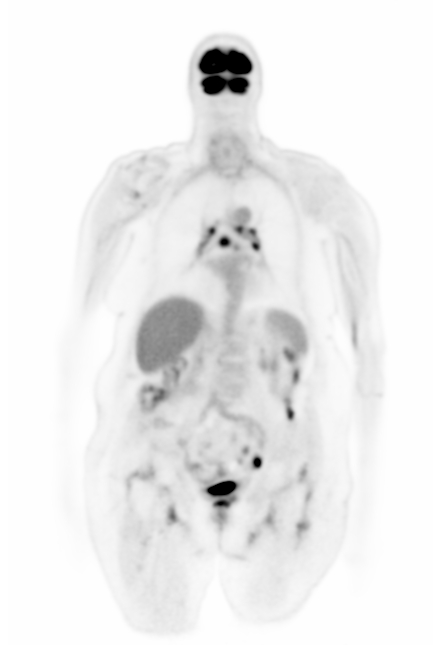} &
				\includegraphics[width=2\tempdima]{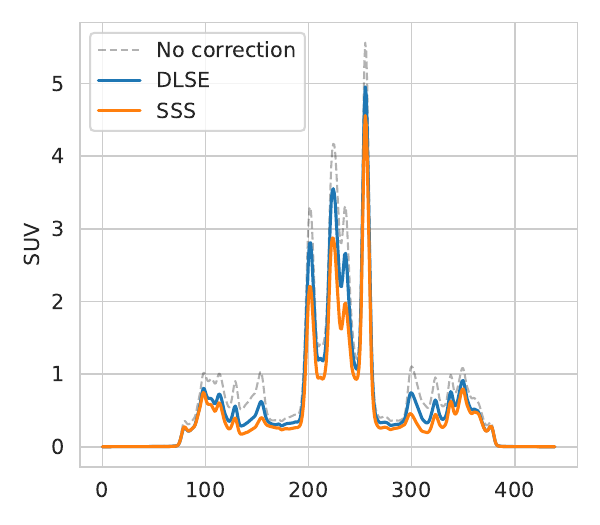} \\
				
				\includegraphics[width=\tempdima]{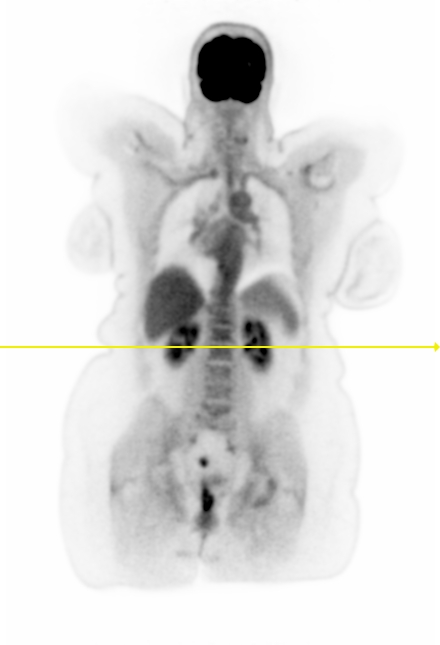} &
				\includegraphics[width=\tempdima]{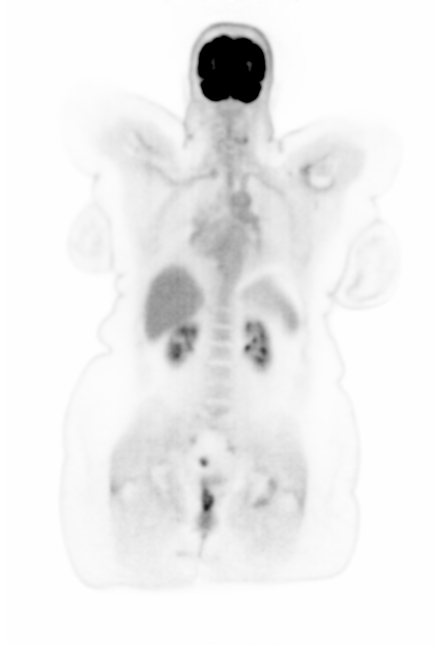} &
				\includegraphics[width=\tempdima]{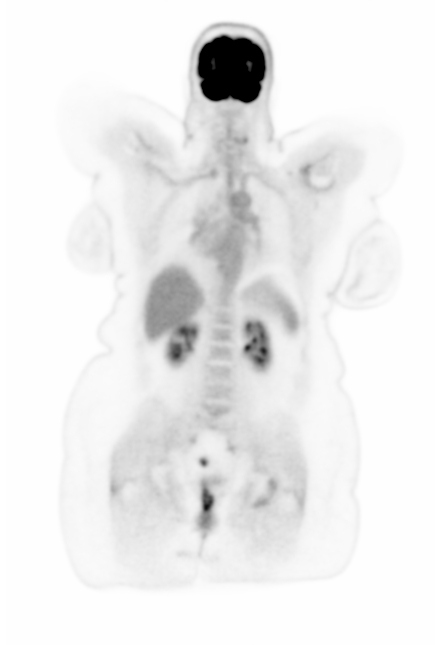} &
				\includegraphics[width=2\tempdima]{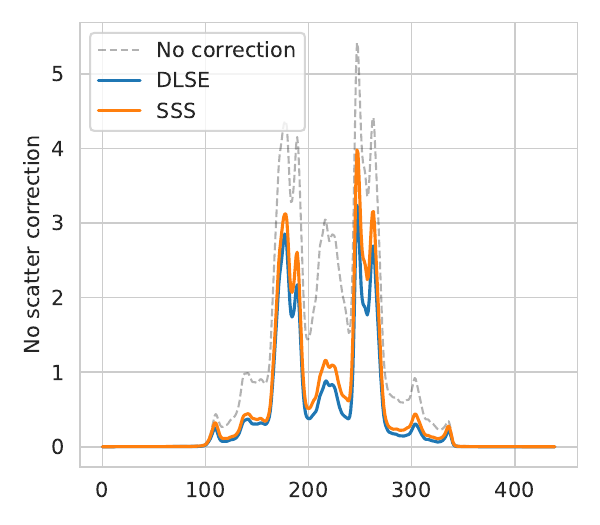} \\
				
		\end{tabular}
	
		\begin{tabular}{ccccc} 
			
			\toprule[\heavyrulewidth]
			& Sex 			& Weight & Dose & Coincidences  \\ 
			\midrule
			1st row	& Female & 52 kgs & 160 MBq & 3.0 billions\\
			2nd row	& Male & 80 kgs & 244 MBq & 4.7 billions\\
			3rd row	& Female & 98 kgs & 297 MBq & 3.8 billions\\
			\bottomrule[\heavyrulewidth]
		\end{tabular}
		
	\caption{
		Clinical \ac{FDG} data reconstructed without scatter correction as well as  with \ac{DLSE} and \ac{SSS} corrections. Profile are shown along the yellow lines drawn of the first column. The same contrast is applied to the three images of a single row.
		}
	\label{fig:images_clinical_FDG}
\end{figure*}

Figure~\ref{fig:images_clinical_FDG} shows three clinical \ac{FDG} dataset examples. The \ac{DLSE}-corrected images are visually very similar to the \ac{SSS}-corrected images. Note that the second \ac{PET} acquisition was performed in  arms-down position while our model was trained in  arms-up position. However this does not seem to affect the results.

Examining the profile lines in the first row, which displays a breast lesion in a female patient, reveals similar patterns to those observed in the simulated images in Figure~\ref{fig:images_simulated_lesions}, with lower activity in the region between the lungs using \ac{SSS} compared to \ac{DLSE}. However, the activities within the lesion are very similar for both \ac{DLSE} and \ac{SSS}. 

In the second row dataset, \ac{DLSE} provides better contrast on the nodules than \ac{SSS}, showing similar activity levels between the nodules but with a higher activity peak.

The last dataset shows kidney structures in a large morphology patient (98~kg). The \ac{SSS}-corrected image exhibits an overall higher activity than the \ac{DLSE}-corrected image. After manual segmentation of the kidneys and automatic segmentation of its structures using the FLAB algorithm \cite{flab_hatt}, the contrast is found to be slightly higher in the \ac{DLSE} corrected \ac{PET} images. The contrast  is 2.3 with \ac{DLSE} correction, compared to 2.2 with \ac{SSS} correction and 1.8 in the uncorrected image.

\begin{figure}
	\centering
	\includegraphics[width=0.7\linewidth]{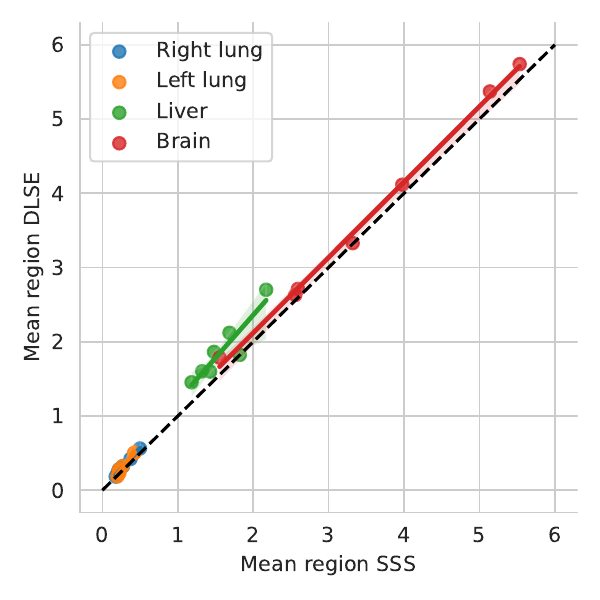}
	\caption{Mean \ac{SUV} correlation between \ac{DLSE} and \ac{SSS}-corrected images within different organs}
	\label{fig:clinical_FDG_region_results}
\end{figure}

Figure~\ref{fig:clinical_FDG_region_results} shows the correlation between activities in \ac{DLSE}- and \ac{SSS}-corrected \ac{PET} images across different organs. Considering all \acp{ROI}, the relationship between the mean activities of both methods is given by the regression function $y=1.03x+0.08$. Activity values are similar in the brain and lung regions, but \ac{DLSE}-corrected images tend to show higher activity levels in the liver compared to \ac{SSS}-corrected images.

\subsubsection{PSMA Datasets}

In this section, we present the results of our \ac{FDG}-trained \ac{DLSE} model applied to \ac{PSMA} to evaluate its ability to generalise to radiotracers not included in the training data.

\begin{figure*}
	
	\settoheight{\tempdima}{\includegraphics[height=0.15\linewidth]{FDG_038_noSC.jpg}}%
	\centering
	
	\begin{tabular}{@{\hspace{-0.22cm}}c@{\hspace{-0.25cm}}c@{\hspace{-0.25cm}}c@{\hspace{0cm}}c@{}} 
		\scriptsize No correction  & \scriptsize DLSE & \scriptsize SSS & \scriptsize Profiles \\

		\includegraphics[width=\tempdima]{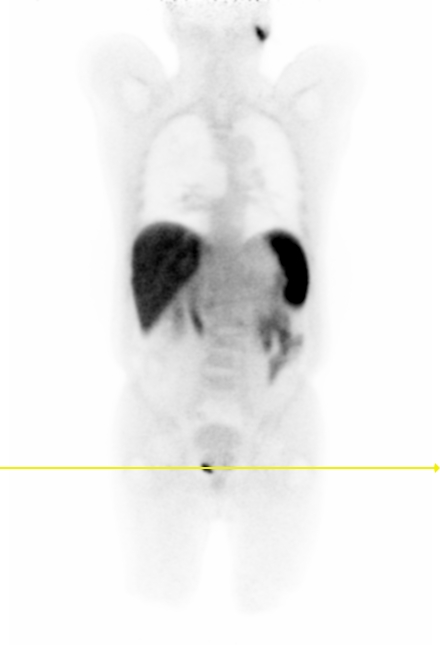} &
		\includegraphics[width=\tempdima]{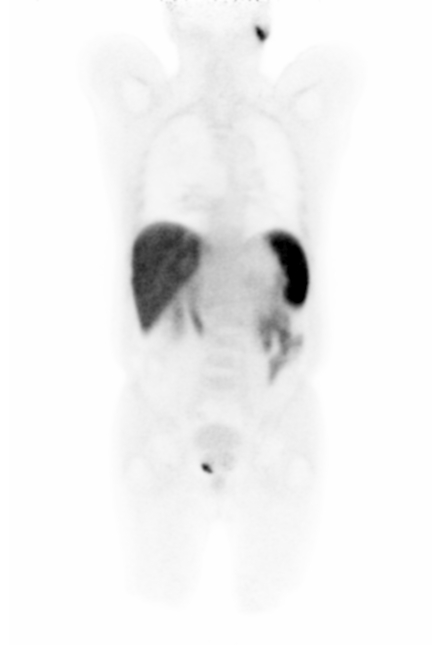} &
		\includegraphics[width=\tempdima]{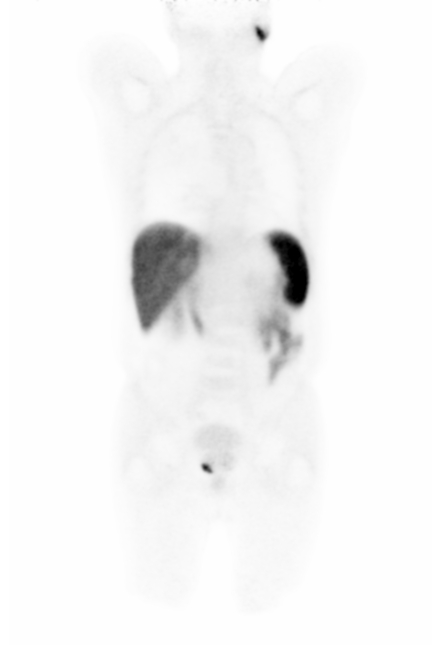} &
		\includegraphics[width=2\tempdima]{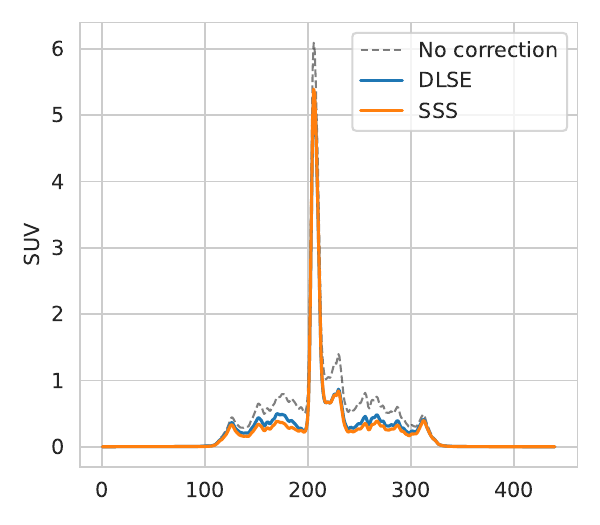} \\

		\includegraphics[width=\tempdima]{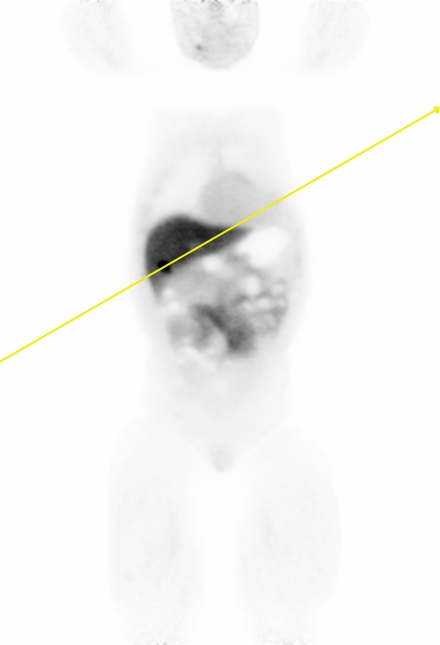} &
		\includegraphics[width=\tempdima]{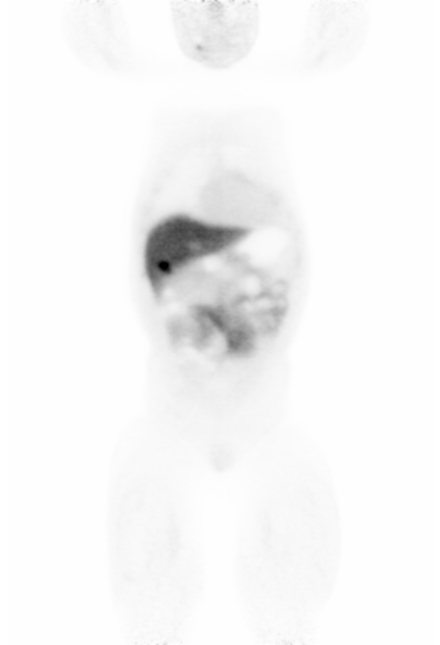} &
		\includegraphics[width=\tempdima]{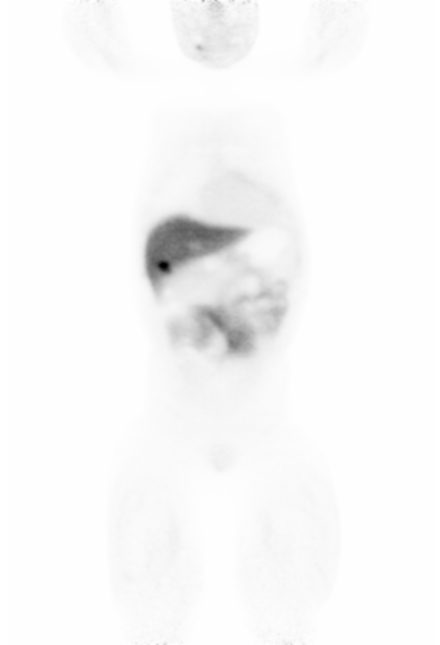} &
		\includegraphics[width=2\tempdima]{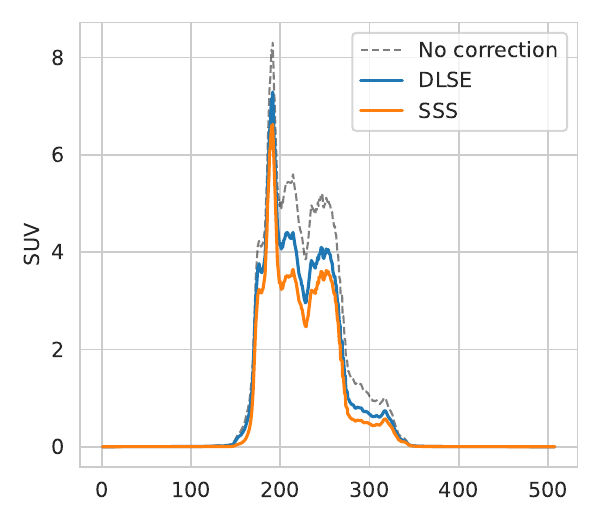} \\
		
	\end{tabular}

		\begin{tabular}{ccccc} 
			
			\toprule[\heavyrulewidth]
			& Sex 			& Weight & Dose & Coincidences  \\ 
			\midrule
			1st row	& Male & 82 kgs & 202 MBq & 743 millions\\
			2nd row	& Male & 73 kgs & 198 MBq & 738 millions\\
			\bottomrule[\heavyrulewidth]
		\end{tabular}
	\caption{
		Clinical \ac{PSMA} data reconstructed without scatter correction as well as with \ac{DLSE} and \ac{SSS} corrections. Profile  are shown along the yellow lines drawn of the first column. The same contrast is applied to the 3 images of a single row.
	}
	\label{fig:images_clinical_PSMA}
\end{figure*}

Figure~\ref{fig:images_clinical_PSMA} shows two clinical \ac{PSMA} dataset examples. The first dataset illustrates a patient with a prostate lesion and shows good agreement between the \ac{DLSE} and \ac{SSS} corrected \ac{PET} images for both the organs and the lesion.

The second patient example features two adjacent liver lesions; one with a higher and one with a lower activiy level compared to the background liver activity. The \ac{DLSE}-corrected image shows higher activity levels in both the liver and the lesions. Based on a \ac{3D} manual segmentation of the liver and the FLAB segmentation of the lesions, the contrast values were found to be similar for \ac{DLSE} and \ac{SSS}, with values of $0.87$ and $0.86$, respectively, for the necrotising lesion, and $1.64$ and $1.67$, respectively, for the active lesion. These values compare with contrasts of $0.92$ and $1.53$, respectively, in the non-scatter-corrected image.

\begin{figure}
	\centering
	\includegraphics[width=0.7\linewidth]{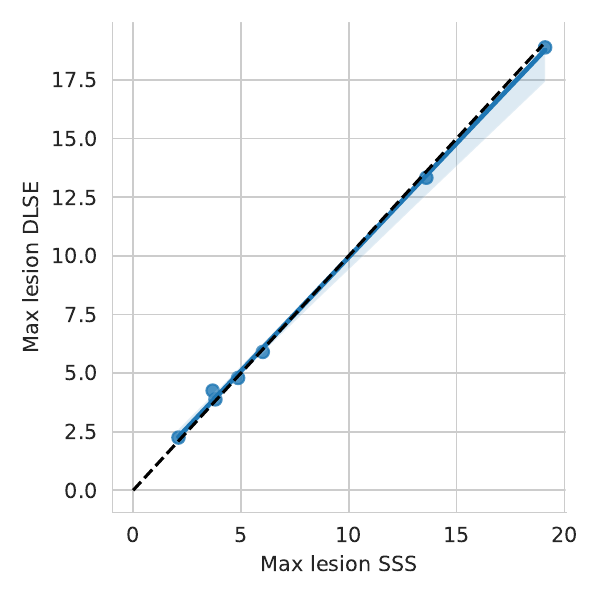}
	\caption{\ac{DLSE} and \ac{SSS}-corrected images \ac{SUVmax} correlation within lesions of \ac{PSMA} dataset}
	\label{fig:clinical_PSMA_lesions_results}
\end{figure}

Figure~\ref{fig:clinical_PSMA_lesions_results} displays the correlation between \ac{DLSE} and \ac{SSS} \ac{SUVmax} values for lesions in the \ac{PSMA} dataset, highlighting good agreement between the two methods, with a relationship defined by the affine function $y=0.97x+0.24$.

\subsection{Computational Speed}

The \ac{DLSE} method was run on an Intel Xeon 3.70~Ghz 10-core CPU and NVIDIA RTX A6000 GPU. The learning step took two hours per epoch resulting in 20 hours for the whole training process. The prediction of a whole \ac{3D} scatter sinogram of 11,559 slices takes 381 seconds (33~ms per sinogram slice).

\section{Discussion}\label{sec:discussion}

This study aims to assess the performance of a previously proposed deep learning scatter estimation method, \ac{DLSE}, using raw \ac{PET} data in the context of \ac{LAFOV} systems. It is worth noting that the training of the proposed \ac{DLSE} approach is entirely based on the use of simulated datasets, hence removing the need for clinical datasets in the training process. The first step involved evaluating the method on simulated images, allowing for comparison with the ground truth in both the sinogram and image domains. In the second step, \ac{DLSE} was tested on 14 clinical datasets, encompassing a range of patient morphologies, tumour sizes, and locations, as well as two different radiopharmaceuticals.

The sinograms produced by \ac{DLSE} from simulated datasets are visually similar to the ground truth and exhibit comparable distributions (Figure~\ref{fig:sino_simulated}). \Ac{DLSE} also demonstrates better scatter quantification than the \ac{SSS} sinogram, particularly for larger phantoms, where \ac{SSS} tends to underestimate scatter levels. This underestimation is more likely in cases of large morphologies that occupy a significant portion of the system's \ac{FOV}, leading to inaccuracies in the tail-fitting scale factor used to account for multiple scatter events. The results shown in Figure~\ref{fig:sinos_simulated_results} confirm that the scatter sinograms produced by \ac{DLSE} are more accurate and less sensitive to phantom size and dose than \ac{SSS}-produced sinograms.

Reconstruction of simulated data corrected with \ac{DLSE} closely resembles the reference image, exhibiting no significant artefacts in various organs or around the lesions (Figure~\ref{fig:images_simulated_lesions}). In addition to demonstrating better robustness to phantom size, the quantitative analysis reveals that \ac{DLSE} is more resilient to variations in injected dose levels, showing lower \ac{NRMSE} disparity compared to \ac{SSS} in the reconstructed \ac{PET} images (Figure~\ref{subfig:images_phantom_dose}). In addition to demonstrating better robustness to phantom size, \ac{DLSE} provides better accuracy than \ac{SSS} on all considered doses. The performance of \ac{DLSE} consistently improves as the patient activity increases but it decreases as the patient size increases (Figure~\ref{subfig:images_phantom_size} and Figure~\ref{subfig:images_phantom_dose}). Furthermore, the analysis of different organ activities (Figure~\ref{subfig:images_simulated_ROI}) indicates an improved accuracy for low-activity regions, such as lungs, as well as for high-activity brain region. \Ac{DLSE} also proved to be more accurate, providing closer contrast recovery than \ac{SSS}, when compared to reference image lesion contrasts (Figure~\ref{subfig:images_phantom_lesions_means}).

The study conducted on clinical \ac{FDG} acquisitions demonstrated consistent results for \ac{DLSE}, producing visually comparable outcomes to \ac{SSS}-corrected images (Figure \ref{fig:images_clinical_FDG}). The method appeared to be robust against significant variations in patient morphology, with weights ranging from 52 to 98 kg. In two of the three cases, \ac{DLSE}-corrected images exhibited slightly higher activity levels, while the \ac{SSS} method showed higher activity in the large morphology patient. This discrepancy may be attributed to inaccuracies in the tail-fitting algorithm used to estimate the scaling factor for multiple scatters, as the tails could potentially be too small for larger morphology patients. In all three examples, the lesion contrasts were found to be greater than those obtained with the \ac{SSS}-based scatter correction.

It is also worth noting that the acquisition duration for these clinical datasets is twice as long as that of the simulated training data (six minutes compared to three minutes). These longer acquisition times naturally result in a higher number of detected coincidences, with a mean of $1.2 \pm 0.55 \times 10^9$ for the simulated dataset, compared to a mean of $3.6 \pm 0.6 \times 10^9$ coincidences for the \ac{FDG} datasets. Consequently, the method is capable of adapting to different statistical levels in the data, which may arise from variations in radiopharmaceutical doses or acquisition durations.

Finally, Figure \ref{fig:images_clinical_PSMA} shows visually similar results between \ac{DLSE} and \ac{SSS}-corrected images. Furthermore, \acp{SUVmax} were found to be similar between the two methods (Figure \ref{fig:clinical_PSMA_lesions_results}). These results suggest that the method could adapt from one radiopharmaceutical to another without the need for retraining.

The prediction of the \ac{3D} scatter sinogram takes about 380 seconds. In comparison, the \ac{SSS} \ac{3D} scatter estimation requires approximately 100 seconds, while the \ac{DSS} can take an average of 5.9 times longer than the \ac{SSS} on systems with a large number of \ac{TOF} bins \cite{watson_DSS}. The speed performance of \ac{DLSE} can be significantly improved by reducing the resolution of the input scatter sinograms. Specifically, the output scatter sinogram is smoothed with a Gaussian filter before being incorporated into the reconstruction. For this study and the resulting qualitative and quantitative sinogram analysis, we opted to maintain the original sinogram dimensions. However, the pre-reconstruction smoothing step could be replaced with sinogram downsampling prior to network training without losing significant information in the final scatter sinogram. Reducing the resolution by a factor of 2 in each dimension would yield a level of detail comparable to that of the Gaussian-smoothed sinogram, and would decrease the scatter estimation duration to less than 50 seconds.

Realistic \ac{MC} \ac{PET} simulations are computationally demanding; however, in the context of a \ac{DL}-based method that utilises simulated data for training, the simulation processes only need to be run once.

\Ac{DLSE} was tested on relatively high-count data. Additional experiments should include low-count \ac{PET} configurations such as ultra low-dose acquisitions or dynamic studies involving short time frames.

In this study, the sinograms were generated using a full-angle acceptance configuration with an \ac{MRD} of 322. Since the method is applied independently to each sinogram slice, it can be readily used with \ac{MRD} 85 data as they represent a subset of the \ac{MRD} 322 sinograms.

The presented work has been applied to non-\ac{TOF} \ac{PET} data. A straightforward approach for adapting the method to \ac{TOF} data, which represents a potential direction for future investigations, involves applying the method to each time bin sinogram.

Finally, we compared \ac{DLSE} method to \ac{SSS} as it is the most common sinogram-based scatter correction method. However, a comparative study with the state-of-the-art image-based strategies would be valuable.

\section{Conclusions}\label{sec:conclusion}

In the present study we assessed the \ac{DLSE} method on a clinical \ac{LAFOV} \ac{PET} system, representing more challenging scatter conditions compared to standard \ac{FOV} \ac{PET} scanners. This method offers the advantage of directly incorporating multiple scatters and oblique planes in the data correction process. Similarly with the performance previously shown on conventional \ac{PET} systems, \ac{DLSE} demonstrated higher accuracy on phantom data, demonstrating better robustness to variations in patient size and injected dose levels in comparison to \ac{SSS}, also providing better lesion contrast recovery.
\ac{DLSE} also yielded high-quality results on clinical data, showing improved lesion contrasts on \ac{FDG} datasets and consistent results on \ac{PSMA} datasets, despite \ac{PSMA} activity distributions not being used during the \ac{DLSE} training process. This study indicates that deep-learning methods applied to raw \ac{PET} data are effective for scatter estimation and correction in \ac{LAFOV} PET systems. Future investigations could focus on \ac{DLSE} generalisation across multiple scanner geometries without requiring retraining.

\section*{Statements \& Declarations}

\paragraph{Acknowledgments} The authors would like to thank Jorge Cabello from Siemens Healthineers for sharing their tool for generating Siemens-format list-mode files from simulation files. 

\paragraph{Funding}

The authors declare that no funds, grants, or other support were received during the preparation of this manuscript.

\paragraph{Competing Interests}

Axel Rominger and Kuangyu Shi are editors of this journal.

\paragraph{Author Contributions}
All authors contributed to the study conception and design. Material preparation, data collection and analysis were performed by Baptiste Laurent. The first draft of the manuscript was written by Baptiste Laurent and Alexandre Bousse. The clinical data were prepared by Kuangyu Shi and Alex Rominger. The data converter used to convert Monte-Carlo simulations output files to Siemens list-mode files format was developed by Jorge Cabello. All authors read and approved the final manuscript.

\paragraph{Data Availability}
The datasets generated during and/or analysed during the current study are available from the corresponding author on reasonable request.

\paragraph{Ethics approval}
This research does not involve human subjects---patient data were used retrospectively.

\bibliography{DLSE_quadra}

\end{document}